\documentclass{article}
\usepackage{latexsym, amssymb, enumerate, amsmath, graphicx, listings}
\usepackage[framed]{mcode}

= 0.5\thickmuskip \arraycolsep = 0.3\arraycolsep

\begin{document}

\title{Entropy Balance and Dispersive Oscillations \\ in Lattice
Boltzmann Models}

\author{Dave Packwood \thanks {Department of Mathematics, University of
Leicester, (dp123@le.ac.uk).}}

\date{}

\maketitle

\begin{abstract}
We conduct an investigation into the dispersive post-shock
oscillations in the entropic lattice-Boltzmann method (ELBM). To
this end we use a root finding algorithm to implement the ELBM which
displays fast cubic convergence and guaranties the proper sign of
dissipation. The resulting simulation on the one-dimensional shock
tube shows no benefit in terms of regularization from using the ELBM
over the standard LBGK method. We also conduct an experiment
investigating of the LBGK method using median filtering at a single
point per time step. Here we observe that significant regularization
can be achieved.
\end{abstract}

\smallskip


\section{Introduction}

The Lattice Boltzmann methods in their original form (see
\cite{Benzi,PolyEquil}) do not guarantee the proper entropy
production and may violate the Second Law. The proper entropy
balance remains up to now a challenging problem for many lattice
Boltzmann models \cite{Nonexist}.

The Entropic lattice Boltzmann method (ELBM) was invented first in
1998 as a tool for construction of single relaxation time lattice
Boltzman models which respect the  $H$-theorem \cite{Htheorem}. For
this purpose, instead of the mirror image with local equilibrium as
reflection center, the entropic involution was proposed, which
preserves the entropy value. Later, we call it the Karlin-Succi
involution \cite{gorban06}. In 2000, it was reported that exact
implementation of the Karlin-Succi involution (which keeps the
entropy balance) significantly regularizes the post-shock dispersive
oscillations \cite{ELBM}. This regularization seems very surprising,
because the entropic lattice BGK (ELBGK) model gives a second-order
approximation to the Navier--Stokes equation (different proofs of
that degree of approximation were given in \cite{PolyEquil} and
\cite{BGJ}), and due to the Godunov theorem \cite{Godunov} linear
second-order finite difference methods have to be non monotonic.

Moreover, Lax \cite{Lax} and Levermore with Liu
\cite{Levermore1996}, demonstrated that these {\it dispersive
oscillations} are unavoidable in classical numerical methods.
Schemes with precise control of entropy production, studied by
Tadmor with Zhong \cite{Tadmoor}, also demonstrated post-shock
oscillations. Of course, there remains some gap between methods with
proven existence of dispersive oscillations, and ELBM. However,
recently, the existence of oscillations in the vicinity of the
shock, at small values of viscosity for ELBM, was reported for
Burgers' equation \cite{Bruce}. In a recent paper \cite{Limiters}
post shock oscillations of  ELBGK were reported too, and no
difference was found between ELBGK and LBGK in that regard.

Nevertheless, absence of dispersive oscillations for ELBGK was
reported many times since 2000. In this paper we answer the
question: does the precise control of entropy production by ELBGK
smooth the post-shock oscillation? The answer is negative. The exact
implementation of the entropic involution does not smooth the
dispersive oscillation (similarly, the exact control of entropy
production does not smooth the post shock oscillation in finite
difference methods \cite{Tadmoor}). Hence, the smoothing effect is
caused by numerical imprecision in calculations of entropic
involution, i.e. in solution of the following transcendental
equation with respect to $\alpha$ ($\alpha \neq 0$):
\begin{equation}\label{eq:entinv}
S(f+\alpha(f^*-f))=S(f),
\end{equation}
where $S$ is entropy, $f$ is a current distribution, and $f^*$ is the
corresponding equilibrium.

In the first part of this paper we discuss a different numerical
implementation of the ELBGK and conduct an investigation into
exactly what stabilization properties it exhibits.

The choice of the method for solution of (\ref{eq:entinv}) should be
very precise, and in Section~\ref{numsol} we describe a cubically
converging root finding algorithm. It is not sufficient to have high
precision when we have average deviation of the current distribution
$f$ from the associated equilibrium $f^*$. For example, for
solutions with shocks, it is usual for the distribution of this
deviation to far from being exponential \cite{Limiters}, and there
appear points with deviation of several orders higher than the
average. Moreover, it is sufficient to smooth a solution at {\it one
point} only. We demonstrate this in the second part of the paper. We
select the lattice site with most nonequilibrium $f$ and regularize
the field of nonequilibrium entropy at this point with 3-point
median filter \cite{Limiters}. As a result, the dispersive
oscillations drastically decrease.

\section{Lattice Boltzmann methods}

The Lattice Boltzmann method arises as a discretization of
Boltzmann's kinetic transport equation
\begin{equation}
\frac{\partial f}{\partial t} + \mathbf{v}\cdot\nabla f = Q^c(f).
\label{eq:BoltzmannTransport}
\end{equation}
The \emph{population function} $f$ describes the distribution of
single particles in the system and the \emph{collision integral}
$Q^c$ their interaction. Altogether (\ref{eq:BoltzmannTransport})
describes the behaviour of the system at the microscopic level. By
selecting a finite number of velocities $\mathbf{v}_i, (i =
1,\ldots,n)$ we create discrete approximation of the kinetic
equation in velocity space. An appropriate choice of the velocities
and time step discretizes space. For a time step of $\delta t = 1$
the lattice can be created by unscaled space shifts of the
velocities, and we get the fully discrete lattice Botzmann gas:
\begin{equation}\label{LBMEQ}
    f_i(x+v_i,t+1) =   f_i(x,t)+Q_i
\end{equation}
where the proper transition from continuous collision integral
$Q^c(f)$ to its fully discrete form $\{Q_i\}$ is assumed. The
simplest and the most common choice for the discrete collision
integral $Q_i$ is the Bhatnagar-Gross-Krook operator with
over-relaxation
\begin{equation}
Q_i = \alpha\beta(f_i^\ast  - f_i).
\label{eq:CollisionIntegral}
\end{equation}
For the standard LBGK method $\alpha = 2 $ and $\beta \in [0, 1]$
(usually, $\beta \in [1/2, 1 ]$)  is the over-relaxation coefficient
used to control viscosity. For $\beta = 1/2$ the collision operator
returns the \emph{local equilibrium} $f_i^\ast$ and $\beta = 1$ (the
\emph{mirror reflection}) returns the collision for a liquid at the
zero viscosity limit. For a viscosity $\nu$ the parameter $\beta$ is
chosen by $\beta = \delta t/(2\nu+ \delta t)$. It should be noted
that a collision integral such as (\ref{eq:CollisionIntegral})
demands prior knowledge of a local equilibrium state for the given
lattice.

A variation on the LBGK is the ELBGK \cite{ELBM}. In this case
$\alpha$ is varied to ensure a constant entropy condition according
to the discrete $H$-theorem. In general the entropy function is
based upon the lattice and cannot always be found explicitly.
However in the case of the simple one dimensional lattice with
velocities $\mathbf{v} = (-c,0,c)$ and corresponding populations
$\mathbf{f} = (f_-,f_0,f_+)$ an explicit Boltzmann style entropy
function is known \cite{LatticeEntropies}:
\begin{equation}
S(\mathbf{f}) = -f_-\log(f_-) - f_0\log(f_0/4) - f_+\log(f_+).
\label{eq:ELBMEntropy}
\end{equation}
With knowledge of such a function $\alpha$ is found as the non-trivial root of the equation
\begin{equation}
S(\mathbf{f}) = S(\mathbf{f} + \alpha(\mathbf{f}^\ast - \mathbf{f})).
\label{eq:ELBMCondition}
\end{equation}
The trivial root $\alpha = 0$ returns the entropy value of the
original populations. ELBGK then finds the non-trivial $\alpha$ such
that (\ref{eq:ELBMCondition}) holds. This version of BGK collision
one calls entropic BGK (or EBGK) collision. Solution of
(\ref{eq:ELBMCondition}) must be found at every time step and
lattice site. Entropic equilibria (also derived from the
$H$-theorem) are always used for ELBGK.

\section{The $H$-theorem for LBMs}

In the continuous case the Boltzmann $H$ the Maxwellian distribution
maximizes entropy and therefore also has zero entropy production. In
the context of lattice Boltzmann methods a discrete form of the
$H$-theorem has been suggested as a way to introduce thermodynamic
control to the system \cite{Htheorem}.

From this perspective the goal is to find an equilibrium state
equivalent to the Maxwellian in the continuum which will similarly
maximize entropy. Before the equilibrium can be found an appropriate
$H$ function must be known for a given lattice. These functions have
been constructed in a lattice dependent fashion in
\cite{LatticeEntropies}, and $H = -S$ with S from
(\ref{eq:ELBMEntropy}) is an example of a $H$ function constructed
in this way.

Using equilbria derived from a $H$ function with entropy
considerations in mind leads to a thermodynamically correct LBM.
This is easy to see in the case of the EBGK collision operator
(\ref{eq:CollisionIntegral}) with explicit local equilibrium. EBGK
collision obvioulsly respect the Second Law (if $\beta \leq 1$), and
simple analysis of entropy dissipation gives the proper evaluation
of viscosity.

ELBGK finds the value of $\alpha$ that with $\beta = 1$ (inviscid
fluid) would give zero entropy production, therefore making the
position of zero entropy production the limit of any relaxation. For
the fixed $\alpha$ used in the LBGK method it remains possible,
particularly for low viscosity fluids, to relax past this point
resulting in negative entropy production, violating the Second Law.

Near to the zero-viscosity limit the LBGK method produces spurious
oscillations around shockwaves. Apart from the thermodynamic
benefits of using ELBGK it has been claimed \cite{ELBM} that ELBGK's
thermodynamic considerations act as a regularizer. This claim seems
to be at odds with other numerical methods which respect the same
thermodynamic laws as ELBGK. For example the results of Tadmoor and
Zhong \cite{Tadmoor} for an entropy correct method display intensive
post-shock oscillations. Furthermore it has been demonstrated
\cite{Lax,Levermore1996} that such dispersive oscillations are
artifacts of the lattice rather than thermodynamic issues. As ELBGK
clearly operates on exactly the same lattice as LBGK and other
finite difference schemes it warrants a deeper investigation into
exactly how it achieves the regularization properties claimed.

\section{Computation of entropic involution} \label{numsol}

In order to investigate the stabilization properties of ELBGK it is
necessary to craft a numerical method capable of finding the
non-trivial root in (\ref{eq:ELBMCondition}). In this section we fix
the population vectors $\mathbf{f}$ and $\mathbf{f}^*$, and are
concerned only with this root finding algorithm. We recast
(\ref{eq:ELBMCondition}) as a function of $\alpha$ only:
\begin{equation}
F(\alpha) = S(\mathbf{f} + \alpha(\mathbf{f}^\ast - \mathbf{f})) - S(\mathbf{f}).
\label{eq:NewCondition}
\end{equation}
In this setting we attempt to find the non-trivial root $r$ of
(\ref{eq:NewCondition}) such that $F(r) = 0$. It should be noted
that as we search for $r$ numerically we should always take care
that the approximation we use is less than $r$ itself. An upper
approximation could result in negative entropy production.

The following theorem gives cubic convergence order for a simple
algorithm for finding the roots of a concave function based on local
quadratic approximations to the target function. Analogously to the
case for Newton iteration, the constant in the estimate is the ratio
of third and first derivatives in the interval of iteration.

\smallskip

\noindent{\bf Theorem.} For a three times continously differentiable
concave entropy function $F(\alpha)$ an iterative root finding
method based on the zeros of a second order Taylor parabola has
cubic convergence sufficiently close to the root.

\smallskip

\noindent {\bf Proof:} Assume that we are operating in a
neighbourhood $r \in N$, in which $F'$ is negative (as well of
course $F''$ is negative). At each iteration the new estimate for
$r$  is the greater root of the parabola $P$, the  second order
Taylor polynomial at the current estimate,
\begin{equation}
P(\alpha) = F(\alpha_n) + (\alpha - \alpha_n)F'(\alpha_n) + (\alpha - \alpha_n)^2\frac{F''(\alpha_n)}{2}.
\label{eq:Parabola}
\end{equation}
The Lagrange remainder form of the error is
\begin{eqnarray*}
F(\alpha)  & = & F(\alpha_n) + (\alpha - \alpha_n)F'(\alpha_n) +
(\alpha - \alpha_n)^2\frac{F''(\alpha_n)}{2}
+ (\alpha - \alpha_n)^3\frac{F'''(\gamma_n)}{6} \\
& = & P(\alpha)+(\alpha - \alpha_n)^3\frac{F'''(\gamma_n)}{6},
\end{eqnarray*}
where $\gamma_n$ lies between $\alpha_n$ and $\alpha$. Evaluating
this at $r$ we see that
\begin{equation*}
|P(r)| \le \frac{1}{6}|\alpha_n - r|^3 \sup_{a \in N} |{F'''(a)}|.
\end{equation*}
Now, using the mean value theorem, for some value $b_n \in [\alpha_{n+1},r]$,
$$
|P(r) | =  |P(\alpha_{n+1})+(r-\alpha_{n+1})P'(b_n)|  \ge
|(r-\alpha_{n+1})| \inf_{b \in N} |(F'(b))|.
$$
Combining the last two equations we see that
\begin{eqnarray*}
|(r-\alpha_{n+1})| & \le & C |\alpha_n - r|^3, \;\; \mbox{where}
\;\; C = \frac{1}{6}\sup_{a \in N}|F'''(a)| \left/ \inf_{b \in
N}|F'(b)| \right. . \; \;\; \blacksquare
\end{eqnarray*}
\smallskip

We use a Newton step to estimate the accuracy of the method at each
iteration:
\begin{equation}
|\alpha_n - r| \approx \left| F(\alpha_n)/F'(\alpha_n) \right|.
\label{eq:Stopping}
\end{equation}
In fact we use a convergence criteria based not solely on $\alpha$
but on $\alpha||\mathbf{f}^* - \mathbf{f}||$, this has the intuitive
appeal that in the case where the populations are close to the local
equilibrium $\Delta S = S(\mathbf{f}^*) - S(\mathbf{f})$ will be
small and a very precise estimate of $\alpha$ is unnecessary. We
have some freedom in the choice of the norm used and we select
between the standard $L_1$ norm and the entropic norm. The entropic
norm is defined as  $$||\mathbf{f}^* - \mathbf{f}||_{{f}^*} = -
((\mathbf{f}^* - \mathbf{f}), \left.D^2
S\right|_{{f}^*}(\mathbf{f}^* - \mathbf{f})),$$ where $\left.D^2
S\right|_{{f}^*}$ is the second differential of entropy at point
$\mathbf{f}^*$, and $(x,y)$ is the standard scalar product.

The final root finding algorithm then is beginning with the LBGK
estimate $x_0 = 2$ to iterate using the roots of successive
parabolas. If this first initiation step produces non-positive
population, then the \emph{positivity rule} \cite{BGJ} could be used
(instead of the mirror image we choose the closest value of $\alpha$
which gives non-negative value of populations). The same
regularization rule might be suggested if there exists no root we
are looking for. In the tests described below, this situation never
arose.

We stop the method at the point,
\begin{equation}
|\alpha_{n} - r| \cdot ||\mathbf{f}^* - \mathbf{f}|| < \epsilon .
\label{eq:ConvergenceTwo}
\end{equation}
To ensure that we use an estimate that is less than the root, at the
point where the method has converged we check the sign of
$F(\alpha_n)$. If $F(\alpha_n) > 0 $ then we have achieved a lower
estimate, if $F(\alpha_n) < 0$ we correct the estimate to the other
side of the root with a double length Newton step,
\begin{equation*}
\alpha_n = \alpha_n - 2\frac{F(\alpha_n)}{F'(\alpha_n)} .
\end{equation*}

At each time step before we begin root finding we eliminate all
sites with $\Delta S  < 10^{-15}$. For these sites we make a simple
LBGK step. At such sites we find that round off error in the
calculation of $F$ by solution of equation (\ref{eq:entinv})  can
result in the root of the parabola becoming imaginary. We note that
in such cases a mirror image given by LBGK is effectively indistinct
from the exact ELBGK collision.

We now experimentally study the convergence of the method. The
convergence of the bisection method is presented for control. For
the bisection method we calculate an initial estimate using the root
of the parabola (\ref{eq:Parabola}) with $\alpha_0 = 2$. Whichever
side of the root this estimate is on, an estimate for the opposite
side can be found using a double length Newton step. We then have
both an upper and lower estimate for the root as required for the
beginning of the bisection method. For this test $\epsilon$ is set
to $10^{-7.5}$. This is the maximum accuracy following the bound on
$\Delta S $ of $10^{-15}$ due to the quadratic nature of $F$.

\begin{figure}
\centering
\includegraphics[width=\textwidth]{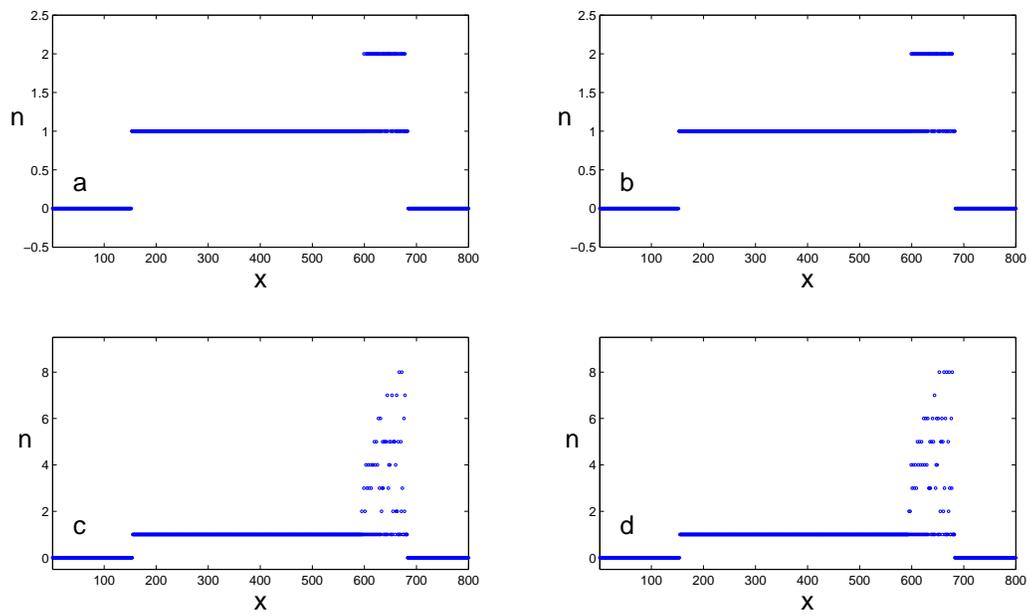}
\caption{Iterations required for convergence to $\epsilon =
10^{-7.5}$ under (\ref{eq:ConvergenceTwo}) using (\textbf{a})
Parabola method with $L_1$ norm; (\textbf{b}) Parabola method with
entropic norm; (\textbf{c}) Bisection method with $L_1$ norm;
(\textbf{d}) Bisection method with entropic norm.
\label{fig:ConvergenceGraph}}
\end{figure}

For shock tube test using 800 lattice sites at the 400th iteration
step (see detailed description in Section~\ref{Tests}),
Fig.~\ref{fig:ConvergenceGraph} shows that the parabola based method
required two iterations, but not more than two, at some points in a
vicinity of shock. In other areas one iteration is sufficient for
the desired accuracy. Across the whole lattice the entropic norm
stipulates a slightly greater number of iterations in both methods.

\section{Shock tube tests}
\label{Tests}

A standard experiment for the testing of LBMs is the one-dimensional
shock tube problem. The lattice velocities used are $\mathbf{v} =
(-1,0,1)$, so that space shifts of the velocities give lattice sites
separated by the unit distance. 800 lattice sites are used and are
initialized with the density distribution
\begin{displaymath}
 \rho(x) =  \left\{
 \begin{array}{ll}
 1, & \;\;\;\;1\leq x \leq 400, \\
 0.5, &\;\;\;\;401 \leq x \leq 800.
 \end{array}
 \right.
\end{displaymath}
Initially all velocities are set to zero. We compare the ELBGK
equipped with the parabola based root finding algorithm using the
entropic norm with the standard LBGK method using both standard
polynomial and entropic equilibria. The polynomial equilibria are
given in \cite{Benzi,PolyEquil}:
\begin{displaymath}
f_-^* = \frac{\rho}{6}\left(1 - 3u + 3u^2\right), \; \; f_0^* =
\frac{2\rho}{3}\left(1 - \frac{3u^2}{2}\right), \; \; f_+^* =
\frac{\rho}{6}\left(1 + 3u + 3u^2\right).
\end{displaymath}
The entropic equilibria also used by the ELBGK are available
explicitly as the maximum of the entropy function
(\ref{eq:ELBMEntropy}),
\begin{displaymath}
f_-^* = \frac{\rho}{6}(-3u - 1 + 2\sqrt{1+3u^2}), \; \; f_0^* =
\frac{2\rho}{3}(2 - \sqrt{1+3u^2}), \; \;  f_+^* = \frac{\rho}{6}(3u
- 1 + 2\sqrt{1+3u^2}).
\end{displaymath}
Now following (\ref{LBMEQ}) the governing equations for the
simulation are
\begin{displaymath}
\begin{split}
 &f_-(x,t+1) = f_-(x+1,t) + \alpha\beta(f_-^*(x+1,t) - f_-(x+1,t)), \\
 &f_0(x,t+1) = f_0(x,t) + \alpha\beta(f_0^*(x,t) - f_0(x,t)), \\
 &f_+(x,t+1) = f_+(x-1,t) + \alpha\beta(f_+^*(x-1,t) - f_+(x-1,t)).
\end{split}
\end{displaymath}
\begin{figure}
\includegraphics[width=\textwidth]{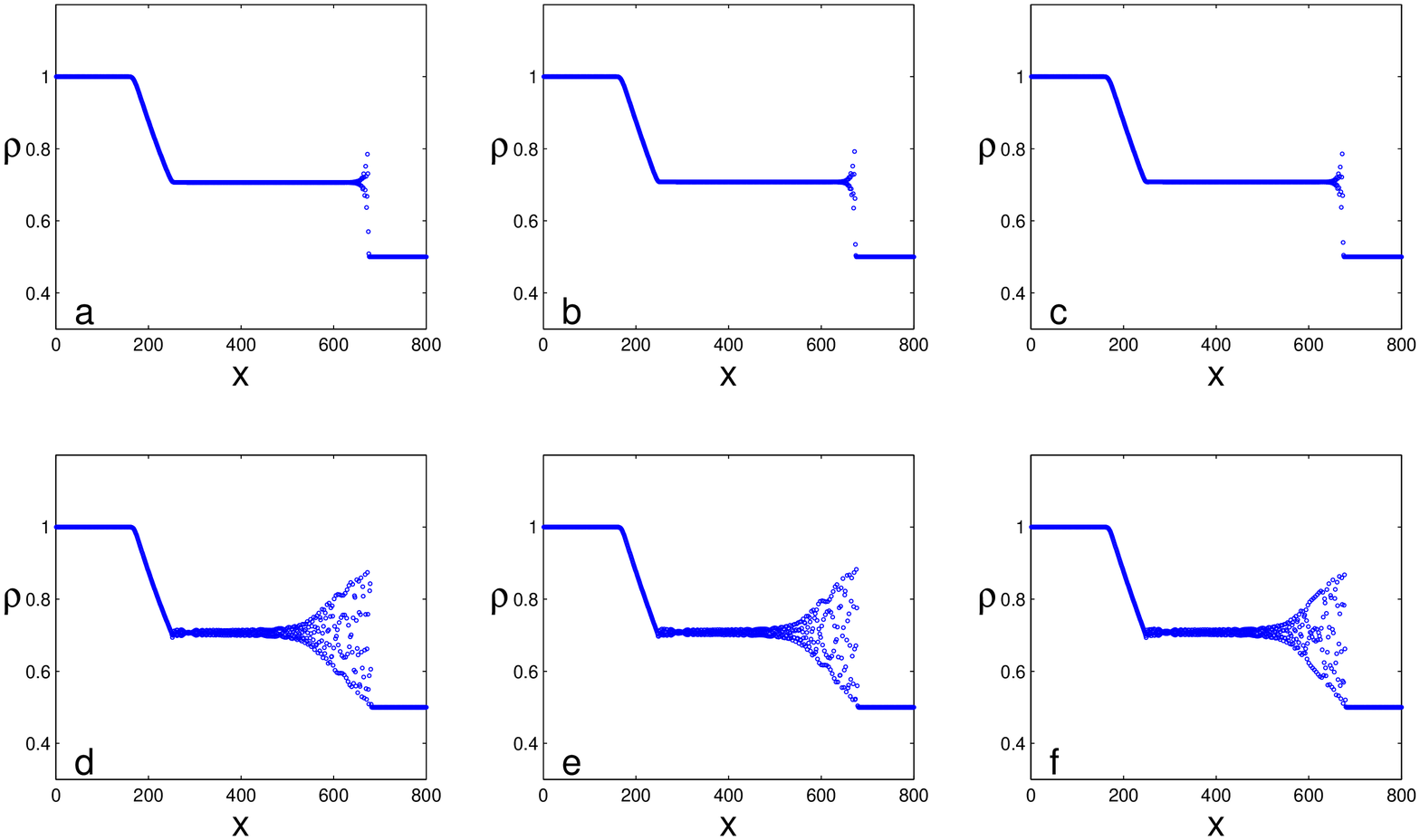}
\caption{Density profile of the simulation of the shock tube problem
following 400 time steps using (\textbf{a}) LBGK with polynomial
equilibria [$\nu = (1/3)\cdot10^{-1}$]; (\textbf{b}) LBGK with
entropic equilibria [$\nu = (1/3)\cdot10^{-1}$]; (\textbf{c}) ELBGK
[$\nu = (1/3)\cdot10^{-1}$]; (\textbf{d}) LBGK with polynomial
equilibria [$\nu = 10^{-9}$]; (\textbf{e}) LBGK with entropic
equilibria [$\nu = 10^{-9}$]; (\textbf{f}) ELBGK [$\nu =
10^{-9}$].\label{fig:ELBMResults}}
\end{figure}
From this experiment we observe no benefit in terms of
regularization in using the ELBGK rather than the standard LBGK
method (Fig. \ref{fig:ELBMResults}). In both the medium and low
viscosity regimes ELBGK fails to supress the spurious oscillations
found in the standard LBGK method.

To explain previous results showing regularization by the ELBGK we
note that in the collision integral (\ref{eq:CollisionIntegral})
that $\alpha$ and $\beta$ are composite. In this sense entropy
production controlled by $\alpha$ and viscosity controlled by
$\beta$ are the same thing. A weak lower approximation to $\alpha$
would lead effectively to addition of entropy at the mostly far from
equilibrium sites and therefore would locally increase viscosity.
This numerical viscosity could, probably, explain the regularization
and smoothing of the shock profile seen in some ELBGK simulations.

\section{One-Point Median Filtering}

\begin{figure}
\includegraphics[width=\textwidth]{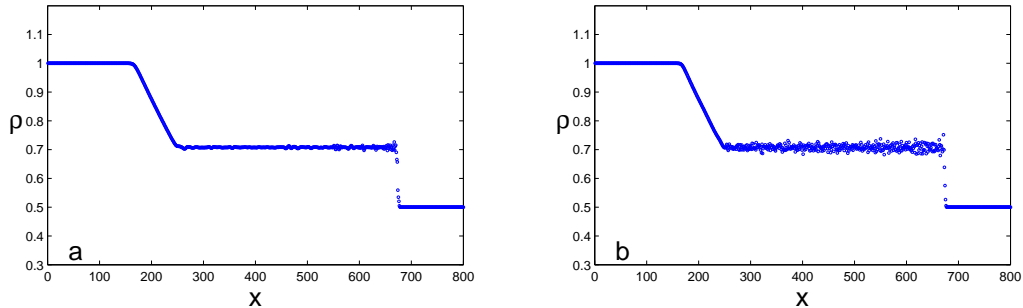}
\caption{Density profile of the simulation of the shock tube problem
following 400 time steps using (\textbf{a}) LBGK with entropic
equilibria and one point median filtering [$\nu =
(1/3)\cdot10^{-1}$]; (\textbf{b}) LBGK with entropic equilibria and
one point median filtering [$\nu = 10^{-9}$].
\label{fig:MedianResults}}
\end{figure}

Finally we consider regularizing the LBGK method using median
filtering at a single point. We follow the prescription detailed in
\cite{Limiters}. First, at each time step, we locate the single
lattice site $x$ with the maximum value of $\Delta S(x)$, and call
this value $\Delta S_x$. Secondly, we find the median value of
$\Delta S$ in the three nearest neighbours of $x$ including itself,
calling this value $\Delta S_{med}$. Now instead of being updated
using the standard BGK over-relaxation this single site is updated
as follows:
\begin{displaymath}
\begin{split}
 &f_-(x,t+1) = f^*_-(x+1,t) + \sqrt{\frac{\Delta S_{med}}{\Delta S_x}}(f_-(x+1,t) - f_-^*(x+1,t)), \\
 &f_0(x,t+1) = f^*_0(x,t) + \sqrt{\frac{\Delta S_{med}}{\Delta S_x}}(f_0(x,t) - f_0^*(x,t)), \\
 &f_+(x,t+1) = f^*_+(x-1,t) + \sqrt{\frac{\Delta S_{med}}{\Delta S_x}}(f_+(x-1,t) - f_+^*(x-1,t)).
\end{split}
\end{displaymath}
We observe that filtering a single point at each time step still
results in  a significant amount of regularization (Fig.
\ref{fig:MedianResults}).

\begin{figure}
\includegraphics[width=\textwidth]{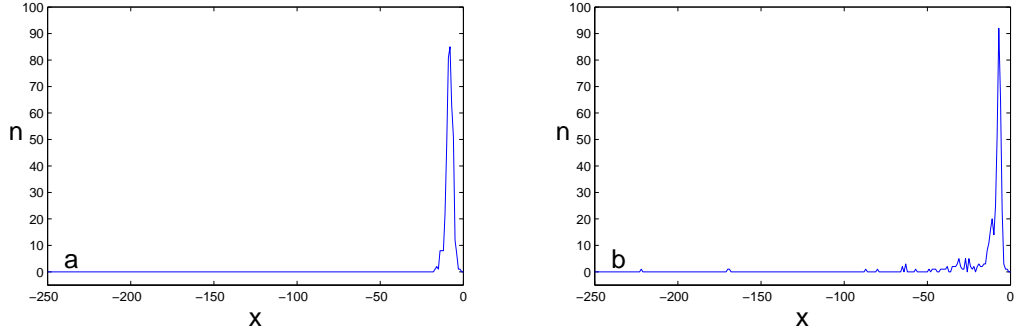}
\caption{Distribution of median filtering sites relative to the
position of the shock following 400 time steps using (\textbf{a})
LBGK with entropic equilibria and one point median filtering [$\nu =
(1/3)\cdot10^{-1}$]; (\textbf{b}) LBGK with entropic equilibria and
one point median filtering [$\nu =
10^{-9}$].\label{fig:MedianPositions}}
\end{figure}

We also examine in each case the lattice site where the filtering is
applied. The zero position is defined as the rightmost lattice site
with $\Delta S > 0$ at each time step and the position of the
filtering is measured relative to this site. The  occurrences at
each relative position are then summed over the experiment. We can
see (Fig. \ref{fig:MedianPositions}) that the majority of filtering
takes place on the shock. However, in the low viscosity case, we
observe that at a small number of time steps the filtered site moves
significantly `behind' the shockwave.

\section{Conclusion}

We present three main conclusions from this study.
\begin{enumerate}
\item We do not find any evidence that
maintaining proper balance of entropy regularize spurious
oscillations the Lattice Boltzmann method. For ELBGK we confirm the
conclusions of Lax \cite{Lax} and Levermore with Liu
\cite{Levermore1996} that {\it dispersive oscillations} are
unavoidable in numerical simulation of shocks.
\item In order to clean up the parasite dispersive oscillations in
the Lattice Boltzmann method it is necessary to filter the entropy
in some way, so as to reduce the extremely-localised incidents of
high non-equilibrium entropy; see \cite{Limiters}. Previously
reported smoothing of shocks must have been via the inadvertent
introduction of numerical dissipation. (Perhaps, this conclusion
could be extended to all known regularisers of LBM, including those
proposed by ourselves in \cite{BGJ}.)
\item For the 1D shock tube, one only needs to filter the entropy at one point per time step
(usually very local to the shock), even at very low viscosity, in
order to effectively eliminate the post-shock oscillation. We can
expect that in 2D and 3D shocks it will be also necessary to filter
nonequilibrium entropy in some local maxima points near the shock
front only. The entropy filtering for non-entropic equilibria is
possible \cite{Limiters} with use of the Kullback--Leibler distance
from current distribution to equilibrium (the relative entropy).
\end{enumerate}

The Matlab code used to produce these results is provided in the appendix.

\newpage
\appendix
\section{Matlab Code}
\begin{lstlisting}
%LBM  Lattice Boltzmann Method 
%    LBM(MC,V,NX,TM,MOV,ESPIL,LIMIT,NORM) solves the shock tube problem 
%    with number of lattice 
%    sites NX with separation of one and viscosity V upto TM time steps of 
%    length one using method 
%    MC:
%     0 - LBGK polynomial equilibria, 
%     1 - LBGK entropic equilibria, 
%     2 - ELBM Newton iterations, 
%     3 - ELBM Bisection method
%    and entropic limiter
%    LIMIT:
%     0 - No limiter
%     1 - Median filtering
%    Output of a movie of the simulation can be controlled with 
%    MOV:
%     0 - No movie,
%     1 - With movie
%    The accuracy of the root finding for ELBM can be controlled with
%    EPSIL in all cases the root found will result in entropy production.
%    Convergence is partly based on ||feq - f||, the choice of the norm is 
%    controlled with 
%    NORM:
%     0 - L1 norm
%     1 - Entropic Norm;

function [rho, mov, convergence, relativeFiltering, alpha] = ...
    LBM(methodChoice,viscosity,nx,timeMax,MOV,ELBMEpsilon,Limiter,Norm)

% Over-relaxation parameter
beta = 1/(2*viscosity+1);
time = 1; %Start Time

rho = LBMInitialise(nx); %Densities at each lattice site are intialised as
%the standard shock tube problem
u = 0; %Velocities at each lattice site are initialised as zero

populations = LBMQuasiEquilibria(rho,u,methodChoice,nx); %Left moving,
%central and right moving populations at each lattice site respectively are 
%initialised as the quasiequilibria

relativeFiltering = zeros(1,251);

if (MOV == 1)
    mov = moviein(timeMax);
    x = 1:nx;
else 
    mov = 0;
end

while time <= timeMax
    populations = LBMPropagate(populations,nx); %Propagate the populations
    %keyboard
    [rho, u] = LBMLatticeParameters(populations); %Density and velocity are
    %calculated using the populations
    
    popequilibriums =  LBMQuasiEquilibria(rho,u,methodChoice,nx); %New 
    % quasiequilibria are found using new density and velocity values
    
    [alpha , convergence] = LBMEntropicParameter(populations, ...
        popequilibriums,methodChoice,nx,ELBMEpsilon,Norm);
   
    % Finding the non trivial root for constant entropy in the ELBM, for
    % normal LBGK  alpha = 0
    
    [limiterSites,shockRelative] = LBMLimiterSites(populations, ...
        popequilibriums,nx,Limiter);
    
    relativeFiltering(250 - shockRelative) = ...
        relativeFiltering(250 - shockRelative) + 1;
    
    populations = LBMCollide(populations,popequilibriums,beta,alpha,nx, ...
        Limiter,limiterSites);

    % Populations are collided with the quasiequilibria
    
    if(~isreal(populations))
        keyboard
        return
    end
    
    if(MOV == 1) % If movie parameter is enabled record a frame
        unlimitedSites = setdiff(1:nx,limiterSites);
        plot(unlimitedSites,rho(unlimitedSites),'.',limiterSites, ...
            rho(limiterSites),'r*'); 
        axis([0 nx 0.3 1.3])
        mov(:,time) = getframe;
    end
    
    time = time + 1; %Increment time
end
\end{lstlisting}
\begin{lstlisting}
%A function to intialise the lattice densities for the shock tube problem
function rho = LBMInitialise(nx)
rho = zeros(1,nx);
half = floor(nx/2);
rho(1:half) = 1;
rho(half+1:nx) = 0.5;
\end{lstlisting}
\begin{lstlisting}
%A function to compute given a current density vector either the polynomial
%or entropic quasiequilbria.
function equilibria = LBMQuasiEquilibria(rho,u,methodChoice,nx)
equilibria = zeros(3,nx); 
if ( methodChoice == 0) %Polynomial Equilibria
    equilibria(1,:) = rho./6.*(1 - 3*u + 3*u.^2);
    equilibria(2,:) = 2.*rho./3.*(1 - 3*u.^2./2);
    equilibria(3,:) = rho./6.*(1 + 3*u + 3*u.^2);
else                    %Entropic Equilibria
    equilibria(1,:) = rho./6.*( -3.*u - 1 + 2*sqrt(1 + 3.*u.^2));
    equilibria(2,:) = 2.*rho./3.*(2 - sqrt(1 + 3.*u.^2));
    equilibria(3,:) = rho./6.*( 3.*u - 1  + 2*sqrt(1 + 3.*u.^2));
end


\end{lstlisting}
\begin{lstlisting}
%A function to propagate the populations
function populations = LBMPropagate(oldpopulations,nx)
populations = oldpopulations;
left = populations(1,1);
right = populations(3,nx);
populations(1,1:nx-1) = populations(1,2:nx);
populations(3,2:nx) = populations(3,1:nx-1);
populations(3,1) = left;
populations(1,nx) = right;
\end{lstlisting}
\begin{lstlisting}
%A function to find density and velocity at each lattice site from the
%populations
function [rho,u] = LBMLatticeParameters(populations)
rho = populations(1,:) + populations(2,:) + populations(3,:);
u = (populations(3,:) - populations(1,:))./rho;
\end{lstlisting}
\begin{lstlisting}
% Function to find the constant entropy parameter for ELBM, in the case of
% LBKG this is simply 2
function [alpha, convergence] = LBMEntropicParameter(populations, ...
    popequilibriums,methodChoice,nx,ELBMEpsilon,Norm)

%Choice of method:
%0 - LBGK polynomial equilibria, 
%1 - LBGK entropic equilibria, 
%2 - ELBM Parabola iterations, 
%3 - ELBM Bisection method

if (methodChoice == 0) 
    alpha = 2*ones(1,nx);
    convergence = zeros(1,nx);
elseif (methodChoice == 1) 
    alpha = 2*ones(1,nx);
    convergence = zeros(1,nx);
elseif (methodChoice == 2)
    convergence = zeros(1,nx);
    alpha = 2*ones(1,nx);
    Sf  = LBMEvaluateS(populations,popequilibriums,zeros(1,nx));
    SEquil =  LBMEvaluateS(popequilibriums,popequilibriums,zeros(1,nx));
    
    remaining = find(abs(SEquil - Sf) > 10^-15);
    popnorm(remaining) = LBMNorms(populations(:,remaining), ...
        popequilibriums(:,remaining),length(remaining),Norm);
          
    iteration = 1;   
    while (isempty(remaining) == 0)
        
        alpha(remaining) = LBMInteriorParabola(populations(:,remaining),...
            popequilibriums(:,remaining),Sf(remaining),alpha(remaining));
        
        Spop =  LBMEvaluateS(populations(:,remaining), ...
            popequilibriums(:,remaining),alpha(remaining));
        Sdiff = LBMEvaluateDiffS(populations(:,remaining), ...
            popequilibriums(:,remaining),alpha(remaining));
            
        deltaAlpha(remaining) = (Spop - Sf(remaining))./Sdiff;
             
        nowdone = find( popnorm(remaining).*abs(deltaAlpha(remaining)) ...
            <= ELBMEpsilon);
        nowremaining =  find( popnorm(remaining).* ...
            abs(deltaAlpha(remaining)) > ELBMEpsilon);
               
        done = remaining(nowdone);     
        remaining = setdiff(remaining,done);  
        
        SDone = LBMEvaluateS(populations(:,done), ...
            popequilibriums(:,done),alpha(done)) - Sf(done);
        negFind = find(SDone < 0);
        negativeEntropy = done(negFind);
        alpha(negativeEntropy) = alpha(negativeEntropy) ... 
            - 2*deltaAlpha(negativeEntropy);

        SDone = LBMEvaluateS(populations(:,done), ...
            popequilibriums(:,done),alpha(done)) - Sf(done);
        if(SDone < 0)
            keyboard
        end
        
        convergence(done) = iteration;
                      
        iteration = iteration + 1;
        if(~isreal(alpha))
            keyboard
        end
        if(iteration > 200)
            keyboard
        end
    end
 
elseif (methodChoice == 3)
    convergence = zeros(1,nx);
    alpha = 2*ones(1,nx);
    Sf  = LBMEvaluateS(populations,popequilibriums,zeros(1,nx));
    SEquil =  LBMEvaluateS(popequilibriums,popequilibriums,zeros(1,nx));
    
    remaining = find(abs(SEquil - Sf) > 10^-14);
    popnorm(remaining) = LBMNorms(populations(:,remaining), ...
        popequilibriums(:,remaining),length(remaining),Norm);
    
    alpha(remaining) = LBMInteriorParabola(populations(:,remaining), ...
        popequilibriums(:,remaining),Sf(remaining),alpha(remaining)) ;
    Spop =  LBMEvaluateS(populations(:,remaining), ...
        popequilibriums(:,remaining),alpha(remaining)) - Sf(remaining);
  
    leftSide = find(Spop > 0);
    rightSide = find(Spop <= 0);
    
    left(remaining(leftSide)) = alpha(remaining(leftSide));
    right(remaining(rightSide))  = alpha(remaining(rightSide));
   
    Spop =  LBMEvaluateS(populations(:,remaining), ...
        popequilibriums(:,remaining),alpha(remaining));
    Sdiff = LBMEvaluateDiffS(populations(:,remaining), ...
        popequilibriums(:,remaining),alpha(remaining));
            
    deltaAlpha(remaining) = (Spop - Sf(remaining))./Sdiff;
    
    right(remaining(leftSide)) = alpha(remaining(leftSide)) ...
        - 2*deltaAlpha(remaining(leftSide));
    left(remaining(rightSide)) = alpha(remaining(rightSide)) ...
        + 2*deltaAlpha(remaining(rightSide));
    
    if(right < left)
        keyboard
    end
    
    iteration = 1;   
    mid = left + (right-left)/2;
    
    
    while (isempty(remaining) == 0)
        mid(remaining) = left(remaining) ...
            + (right(remaining)-left(remaining))/2;
       
       
        Sleft = S(populations(:,remaining), ...
            popequilibriums(:,remaining),left(remaining));
        Sright = LBMEvaluateS(populations(:,remaining), ...
            popequilibriums(:,remaining),right(remaining));
        Smid = LBMEvaluateS(populations(:,remaining), ...
            popequilibriums(:,remaining),mid(remaining));
        
        nowdone = find( popnorm(remaining).*(right(remaining) ...
            - left(remaining)) <= ELBMEpsilon);  
        nowremaining =  find( popnorm(remaining).*(right(remaining) ...
            - left(remaining)) > ELBMEpsilon);  
        done = remaining(nowdone);
        remaining = setdiff(remaining,done);
        alpha(done) = left(done);
        convergence(done) = iteration;
           
        for itr = 1:length(remaining)
            if (Smid(nowremaining(itr)) > Sf(remaining(itr)))
                left(remaining(itr)) = mid(remaining(itr));
            else
                right(remaining(itr)) = mid(remaining(itr));
            end
            
        end        
        iteration = iteration + 1;
        if(~isreal(alpha))
            keyboard
        end
        if(iteration > 100)
        keyboard
        end
    end
end
\end{lstlisting}
\begin{lstlisting}
% A function to evaluate the entropy.
function S = LBMEvaluateS(populations,popequilibriums,alpha)
alphapop = populations + (ones(3,1)*alpha).*(popequilibriums ...
    - populations);
S = -alphapop(1,:).*log(alphapop(1,:)) ...
    - alphapop(2,:).*log(alphapop(2,:)./4) ...
    - alphapop(3,:).*log(alphapop(3,:));
\end{lstlisting}
\begin{lstlisting}
% A function to implement the norms necessary to measure convergence of the
% root finding
function norms = LBMNorms(populations,popequilibriums,nx,nChoice);
norms = zeros(1,nx);
if (nChoice == 0)
    for j = 1:nx
        norms(j) = norm(populations(:,j) - popequilibriums(:,j),1);
    end
elseif(nChoice == 1)
    for j = 1:nx
        poptemp = (populations(1,j) ...
            - popequilibriums(1,j)).^2./popequilibriums(1,j);
        poptemp = poptemp + (populations(2,j) ...
            - popequilibriums(2,j)).^2./popequilibriums(2,j);
        poptemp = poptemp + (populations(3,j) ...
            - popequilibriums(3,j)).^2./popequilibriums(3,j);
        norms(j) = sqrt(poptemp);
    end
end
\end{lstlisting}
\begin{lstlisting}
% A function to find an interior approximation to the root of the entropy
% parabola
function intPab = LBMInteriorParabola(populations,popequilibriums, ...
    STarget,alpha)
SAlphaZero = LBMEvaluateS(populations,popequilibriums,alpha);
SDashAlphaZero = LBMEvaluateDiffS(populations,popequilibriums,alpha);
SDash2AlphaZero = LBMEvaluateDiff2S(populations,popequilibriums,alpha);
intPab = [];
for j = 1:length(alpha)
    intPab(j) = max(roots([0.5.*SDash2AlphaZero(j) SDashAlphaZero(j) ...
        SAlphaZero(j) - STarget(j)])) + alpha(j);
end
\end{lstlisting}
\begin{lstlisting}
% A function to find the first derivative of the entropy
function DiffS = LBMEvaluateDiffS(populations,popequilibriums,alpha)
alphapop = populations + (ones(3,1)*alpha).*(popequilibriums ...
    - populations);
partDiff = popequilibriums  - populations;
DiffS = -partDiff(1,:).*(log(alphapop(1,:)) + 1) ...
    -partDiff(2,:).*(log(alphapop(2,:)./4) + 1) ...
    -partDiff(3,:).*(log(alphapop(3,:)) + 1);
\end{lstlisting}
\begin{lstlisting}
% A function to find the second derivative of the entropy.
function Diff2S = LBMEvaluateDiff2S(populations,popequilibriums,alpha)
alphapop = populations + (ones(3,1)*alpha).*(popequilibriums - populations);
partDiff = popequilibriums  - populations;
Diff2S = - partDiff(1,:).^2./(alphapop(1,:)) ...
    - partDiff(2,:).^2./(alphapop(2,:)) ...
    - partDiff(3,:).^2./(alphapop(3,:));
\end{lstlisting}
\begin{lstlisting}
% A function to detect an appropriate lattice site for limiting and to
% measure it's position relative to the leading edge of the shock.
function [Sites,shockRelative] = LBMLimiterSites(populations,...
    popequilibriums,nx,Limiter)
shockRelative = 0;
if (Limiter == 0) 
    Sites = [];
elseif (Limiter == 1)
    Sf = LBMEvaluateS(populations,popequilibriums,zeros(1,nx));
    Sfeq = LBMEvaluateS(popequilibriums,popequilibriums,zeros(1,nx));
    DeltaS = Sfeq - Sf;
    shockPos = max(find(DeltaS > 10^-15));
    [M,I] = max(DeltaS);
    Sites = I;
    shockRelative = shockPos - Sites;
end
\end{lstlisting}
\begin{lstlisting}
%Function for simple BGK collision of populations with quasiequilibria with
% a limiter applied if necessary
function populations = LBMCollide(oldpopulations,popequilibriums,beta, ...
    alpha,nx,Limiter,limiterSites)
unlimitedSites = setdiff(1:nx,limiterSites);
populations(:,unlimitedSites) = oldpopulations(:,unlimitedSites) ...
    + beta.*(ones(3,1)*alpha(unlimitedSites)) ...
    .*( popequilibriums(:,unlimitedSites) ...
    - oldpopulations(:,unlimitedSites) );
if (Limiter == 1)
    Sf = LBMEvaluateS(oldpopulations(:,(limiterSites - 1): ...
        (limiterSites + 1)),popequilibriums(:,(limiterSites - 1): ...
        (limiterSites + 1)),zeros(1,3));
    Sfeq = LBMEvaluateS(popequilibriums(:,(limiterSites - 1): ...
        (limiterSites + 1)),popequilibriums(:,(limiterSites - 1): ...
        (limiterSites + 1)),zeros(1,3));
    deltaS = Sfeq - Sf;
    Smed = median(deltaS);
    coeff = sqrt(Smed/deltaS(2));
    populations(:,limiterSites) = popequilibriums(:,limiterSites) ...
        + coeff.*(oldpopulations(:,limiterSites) ...
        - popequilibriums(:,limiterSites));
end
\end{lstlisting}

\end{document}